\documentclass[5p]{elsarticle}
\usepackage{graphicx}
\usepackage{citesort}
\usepackage{natbib}
\biboptions{sort&compress}

\title{MOVPE growth of N-polar AlN on 4H-SiC: effect of substrate miscut on layer quality}

\author[ELE]{J. Lemettinen\corref{cor1}}
\ead{jori.lemettinen@aalto.fi}
\cortext[cor1]{Corresponding author}

\author[Tsukuba,MIT]{H. Okumura}

\author[ELE]{I. Kim}

\author[ELE]{C. Kauppinen}

\author[MIT]{T. Palacios}

\author[ELE]{S. Suihkonen}

\address[ELE]{Department of Electronics and Nanoengineering, Aalto University, P.O BOX 13500, FIN-00076 AALTO, Finland}

\address[Tsukuba]{Faculty of pure and applied science, University of Tsukuba, Tsukuba 305-8573 Japan}  

\address[MIT]{Department of Electrical Engineering and Computer Science, Massachusetts Institute of Technology, Cambridge, MA 02139 USA}

\begin{document}

\begin{abstract}

We present the effect of miscut angle of SiC substrates on N-polar AlN growth. The N-polar AlN layers were grown on C-face 4H-SiC substrates with a miscut towards $<\bar{1}100>$ by metal-organic vapor phase epitaxy (MOVPE). The optimal V/III ratios for high-quality AlN growth on 1$^{\circ}$ and 4$^{\circ}$ miscut substrates were found to be 20000 and 1000, respectively. MOVPE grown N-polar AlN layer without hexagonal hillocks or step bunching was achieved using a 4H-SiC substrate with an intentional miscut of 1$^{\circ}$ towards $<\bar{1}100>$. The 200-nm-thick AlN layer exhibited X-ray rocking curve full width half maximums of 203 arcsec and 389 arcsec for (002) and (102) reflections, respectively. The root mean square roughness was 0.43~nm for a 2~$\mu m \times 2~\mu m$ atomic force microscope scan.

\end{abstract}


\maketitle

\section{Introduction}

III-nitride (III-N) semiconductors have received much attention for optical and electrical applications. AlN holds much promise for high-temperature and high-power applications due to its high critical electric field, wide band-gap energy and high thermal conductivity \cite{generalgan,AlNFET}. An AlN-channel field-effect transistor has been demonstrated recently \cite{AlNFET}. However, there are challenges in the charge carrier mobility and contact resistances. These limitations could be overcome by employing nitrogen-polar AlN-based heterostructures.

Wurtzite structure of III-Ns has two types of polar planes, metal-polar (001) and nitrogen-polar (00$\bar{1}$). Generally, III-N high-electron mobility transistors (HEMTs) employ a metal polar AlGaN/GaN structure, which has many merits in crystal growth, i.e., smooth surface, high crystalline quality and low impurity concentrations. However, the performance of metal-polar HEMTs is limited by high contact resistances to the top AlGaN layer and by large leakage current through the GaN buffer layer. In contrast N-polar HEMTs have a reversed structure of GaN/AlGaN layers. The top GaN layer offers low contact resistances \cite{NHEMTreview}. The AlGaN buffer layers with higher band-gap act as a natural back barrier and provide a low leakage current. For N-polar III-Ns HEMTs, high-Al content AlGaN buffer layers can reduce leakage current and induce high carrier concentrations. Thus, N-polar high-Al content AlGaN and AlN growth would be required for further high performance of III-Ns HEMTs.

N-polar AlN buffer layers would be the ideal transistor platform. However, the difficulties of N-polar III-Ns growth are limiting the performance of N-polar III-Ns HEMTs. N-polar III-Ns growth suffers from a high density of threading dislocations and large surface roughness compared to metal polar III-Ns \cite{Nreview,miscut2stepbunch,NGaNSi,1Nnative,NpolarmedNH3,NpolarmedNH3_2,NpolarmedNH3_3,NpolarhighNH3,NpolarhighNH3_2}.

In N-polar III-Ns growth, adatoms have a small diffusion length on surfaces due to high binding energy, limiting adatom transport to kink and step sites and promoting island formation on surface \cite{bindingenergies,Diffusion}. N-polar III-Ns hetero-epitaxial growth has multi-micron hexagonal hillocks formed by inversion domains \cite{microhillocks}. The inversion domains are attributed to the oxidation of substrate or the accumulation of group III atoms at low V/III ratios \cite{1Nnative,NpolarmedNH3_3}. Both surface islands and inversion domains increase surface roughness, making device fabrication unfeasible.

In metalorganic vapor phase epitaxy (MOVPE) growth of N-polar III-Ns, island formations are suppressed using an intentional miscut substrate because the high density of steps in high off-angle substrates enhance a step-flow growth mode \cite{NpolarhighNH3,suppressHillocks,miscut2stepbunch}. However, the growth on high off-angle substrates causes step bunching phenomena, leading to macro steps \cite{miscut2stepbunch,NpolarhighNH3}. 
The step height increases with increasing layer thickness. The direct deposition of high-temperature (Al)GaN layers on an AlN nucleation layer causes macro-steps on the layer surface \cite{LTGaN}. In contrast, step bunching in N-polar (Al)GaN growth is mitigated by nucleating a lower temperature (Al)GaN layer on a thin AlN nucleation layer \cite{LTGaN,NHEMTlowTGaN2}. N-polar III-N devices using low-temperature nucleation layers have been demonstrated \cite{NHEMTlowTGaN1,NHEMTlowTGaN2}. These reports control the growth temperature of N-polar III-Ns at a fixed off angle of substrates. In N-polar AlN growth, there is room to improve the growth mode and crystalline quality by changing the off angle of substrates. 

In a previous study, we have shown that the formation of hexagonal hillocks can be suppressed in N-polar AlN growth by using a low growth rate of 0.1~$\mu m/h$ together with an increased temperature of 1150$^{\circ}C$ \cite{JoriPolarity}. In this paper we study the effect of miscut angle of C-face 4H-SiC substrate on MOVPE growth of N-polar AlN under various growth conditions.

\section{Experiment}

200-nm-thick N-polar AlN layers were grown on C-face of 4H-SiC substrates by MOVPE. The substrates had an intentional miscut of either 4$^{\circ}$ or 1$^{\circ}$ towards $<\bar{1}100>$ (M-plane). The SiC surfaces were cleaned \textit{in-situ} for 20 min at nominal substrate surface temperature of 1180$^{\circ}C$ in hydrogen ($H_{2}$) ambient at 150~mbar pressure prior to AlN growth. The substrate temperature was monitored using an emissivity corrected \textit{in-situ} pyrometer. All temperature values in this work refer to pyrometer estimated substrate temperature. The MOVPE reactor had a close coupled showerhead configuration. The gap between the showerhead and the susceptor was 11~mm, the default value for this reactor. Trimethylaluminium (TMAl) and ammonia ($NH_{3}$) were used as precursors for aluminum and nitrogen, respectively. AlN was grown at 1150$^{\circ}C$ in pressure of 50~mbar with $H_{2}$ carrier gas. The reactor was cleaned between growth runs using high temperature $H_{2}$ and $NH_{3}$ baking steps.

Three-axis high resolution X-ray diffraction (HR-XRD) measurements were performed to assess the crystalline quality of the AlN layers. Rocking curves around symmetrical (002) and skew-symmetrical (102) reflections were recorded. Full width half maximums (FWHMs) for the diffraction peaks were calculated using least-squares fitting of a Lorentzian line-shape. The setup consisted of an X-ray mirror, a 4 $\times$ Ge (220) monochromator and an analyzer crystal. X-ray wavelength of copper $K_{\alpha1}$ emission line was used. Scanning electron microscope (SEM) images were observed in order to assess micrometer-sized surface structures, especially hexagonal hillocks. The surface roughness of the AlN layers was determined by atomic force microscopy (AFM).

\section{Results and discussion}

\begin{figure}[htb]
\begin{center}
\includegraphics[width=0.49\linewidth]{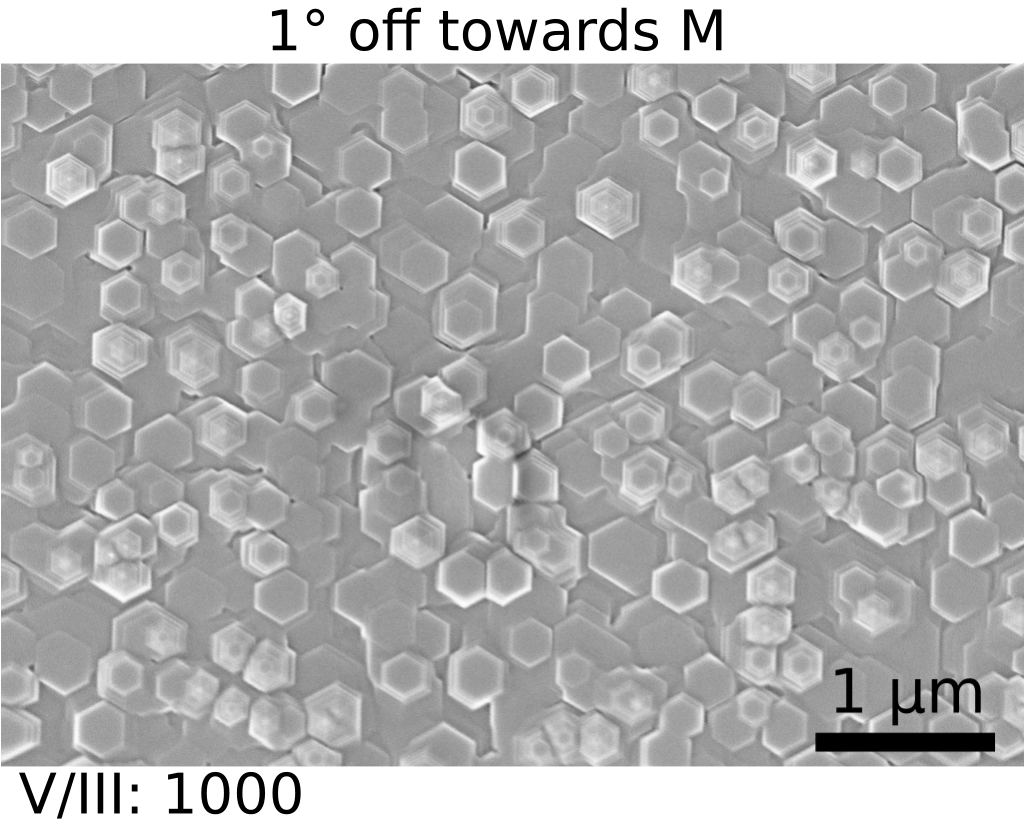}
\includegraphics[width=0.49\linewidth]{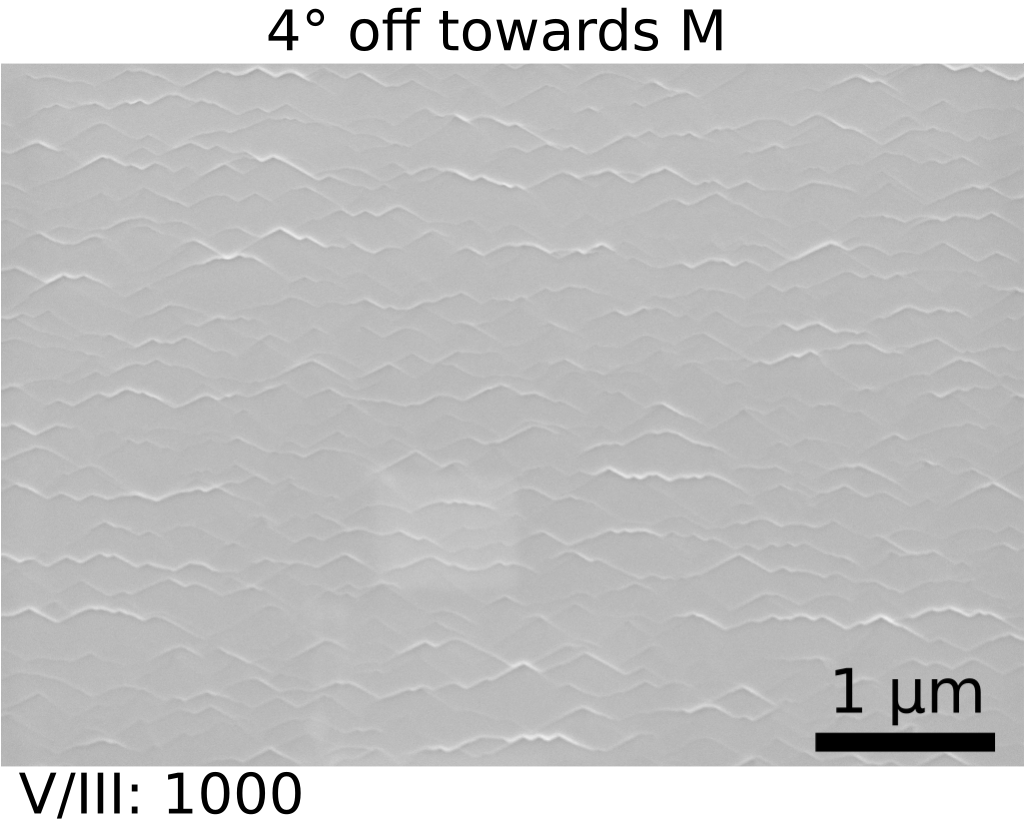}
\\
\includegraphics[width=0.49\linewidth]{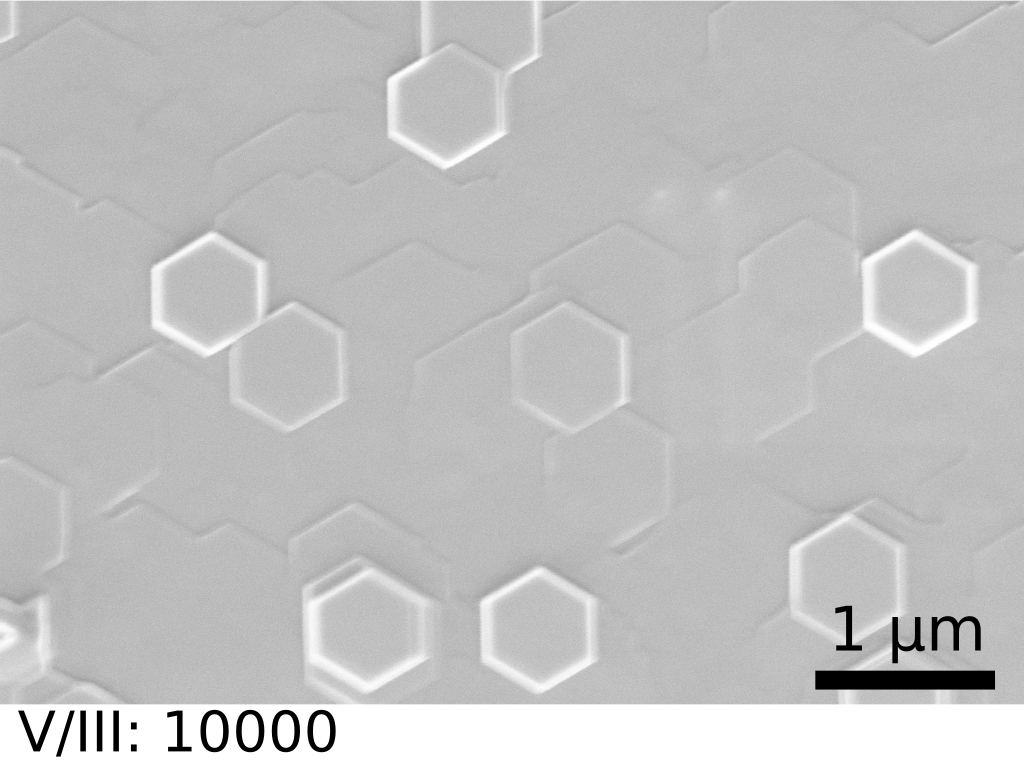}
\includegraphics[width=0.49\linewidth]{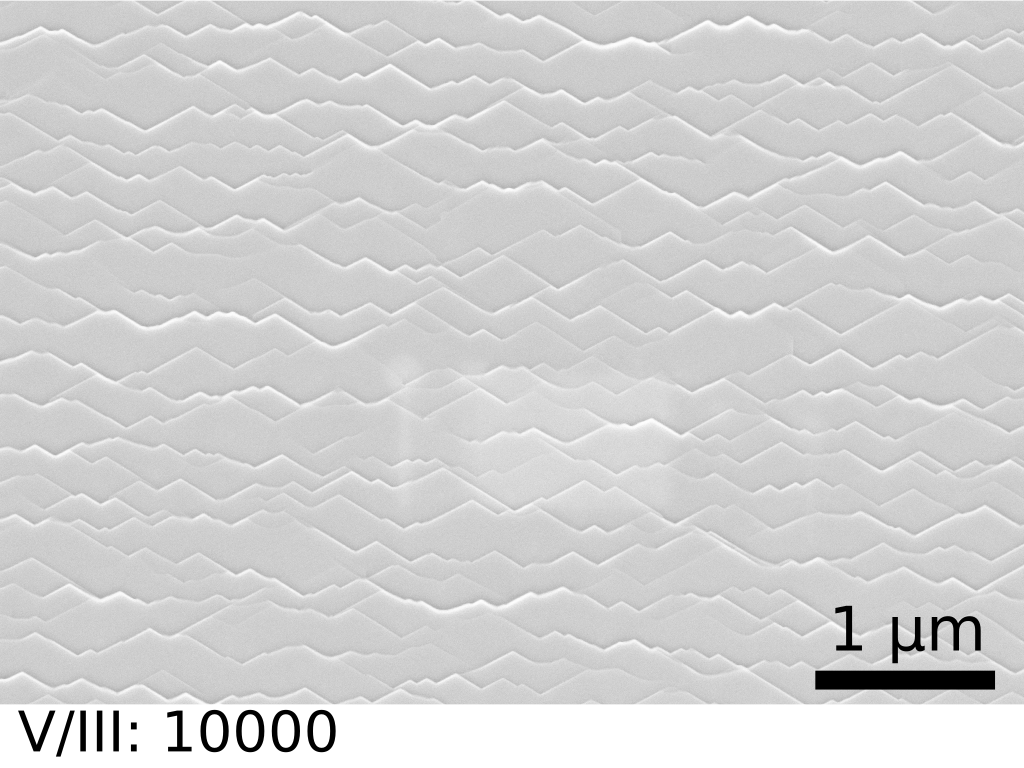}
\\
\includegraphics[width=0.49\linewidth]{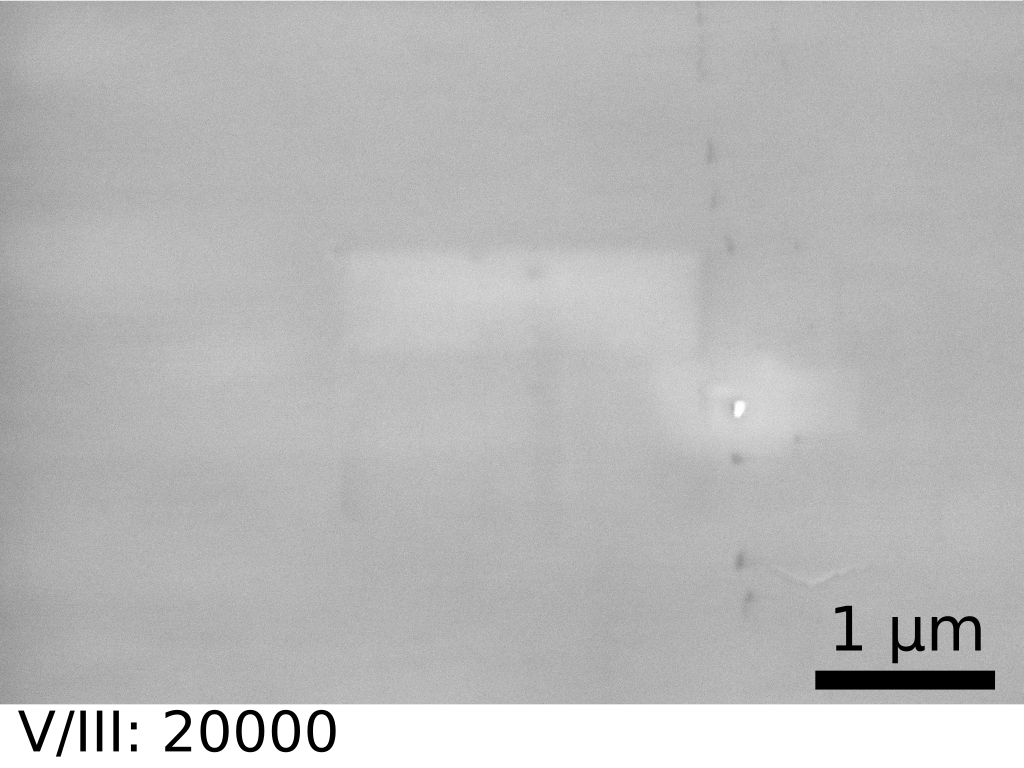}
\includegraphics[width=0.49\linewidth]{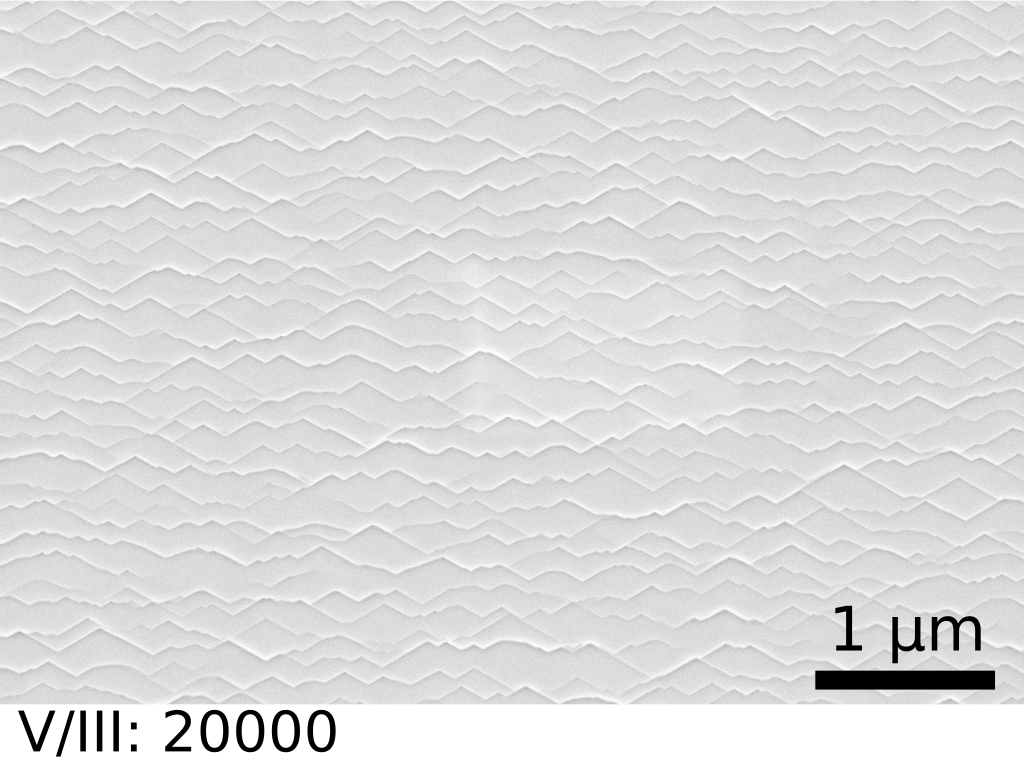}
\\
\includegraphics[width=0.49\linewidth]{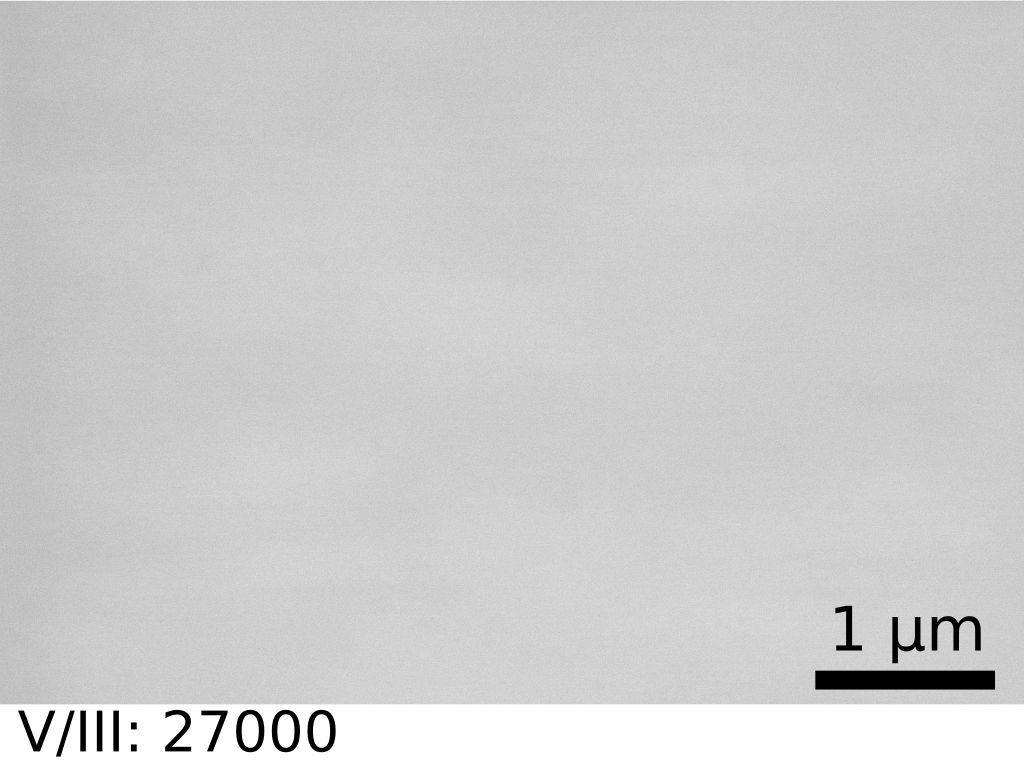}
\includegraphics[width=0.49\linewidth]{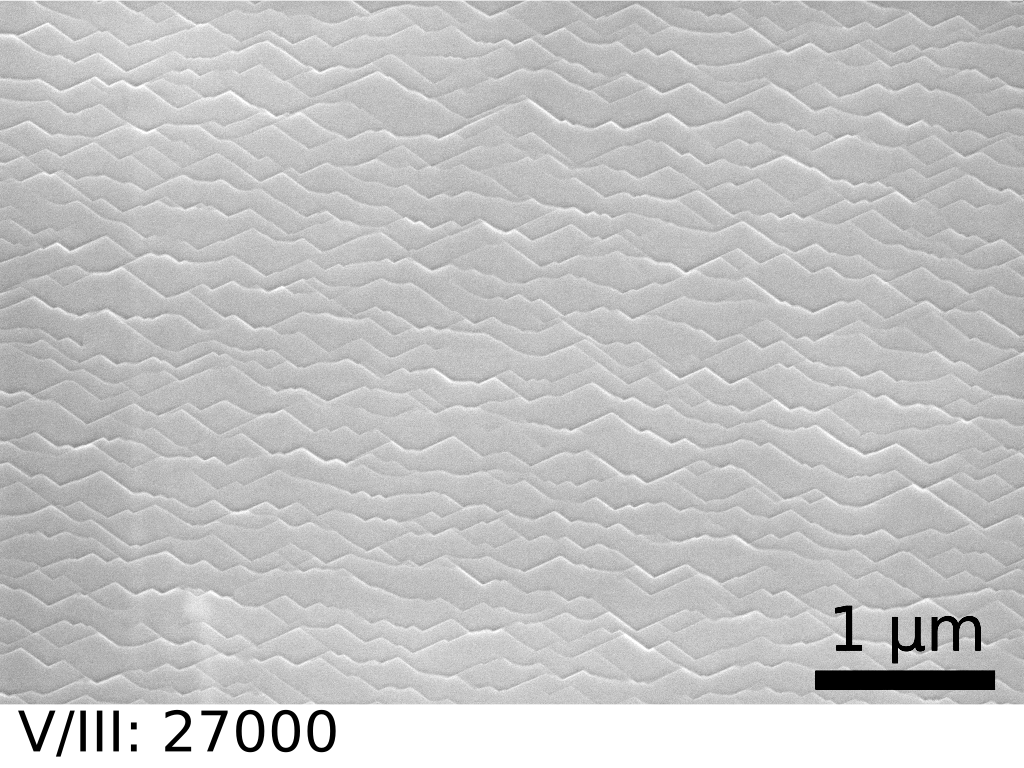}

\caption{SEM images of 200-nm-thick N-polar AlN layers grown on 4H-SiC substrates with either 4$^{\circ}$ or 1$^{\circ}$ miscut towards $[\bar{1}100]$ (M-plane) at 1150$^{\circ}C$ nominal substrate surface temperature with varying V/III-ratio.}
\label{SEM}
\end{center}
\end{figure}

Figure \ref{SEM} shows SEM images of N-polar AlN layers with 200 nm thickness. A series over a wide V/III range from 1000 to 27000 was grown to investigate the effect on N-polar growth in conjugation with the substrate miscut. The growth rate was maintained at 0.1~$\mu m/h$ throughout the series by altering the precursor mass flows to limit the effect of the growth rate on crystalline quality. No multi-micrometer hexagonal hillocks caused by inversion domains were observed in these layers \cite{microhillocks,1Nnative}. All AlN layers grown on the 4$^{\circ}$ miscut substrate exhibited a step flow growth mode. We consider that the step bunching is caused by anisotropy in the terrace step adatom capture probabilities \cite{transition}. The anisotropy in capture probability leads to differences in step lateral advancement rates and can cause step-bunching. The surface roughness increases with increasing layer thickness, indicating that the step bunching effect is not self limiting \cite{NpolarmedNH3_3}.

V/III-ratio has a profound effect on surface morphology of the AlN layers grown on the 1$^{\circ}$ miscut subsrate. Inversion domains have been suggested to form on oxidized regions of the substrate or accumulation of Al-atoms at low V/III ratios \cite{1Nnative,NpolarmedNH3_3}. The AlN layer grown using the V/III-ratio of 1000 exhibits a high density of sub-micrometer hexagonal hillocks. We suggest that the sub-micrometer hillocks result from the enhancement of island formation on the surface terraces since no hillocks are observed on 4$^{\circ}$ miscut subsrates. The dependence of surface kinetics on vapor supersaturation has been used to explain the surface morphology of Al-polar AlN \cite{surface-kinetics-AlN}. Decreasing the substrate miscut increases surface supersaturation, and thus, can promote island formation on the surface terraces if critical supersaturation is reached \cite{surface-kinetics-AlN}.
 
Increasing the V/III-ratio to 10000 significantly reduces the density of hexagonal hillocks and increases the hillock size. The high V/III ratio decreases the surface migration of Al-adatoms, promoting the incorporation of adatoms into the surface kink sites and reducing island formation probability on the surface terraces \cite{NpolarhighNH3,miscut2stepbunch,VIIIdiffusion}. Increasing the V/III-ratio to 20000 suppresses the formation of hexagonal hillocks entirely. Further increase of V/III-ratio to 27000 produces an identical surface morphology. Increasing the V/III-ratio in this growth window shows the elimination of hillocks and the onset of step-flow growth which could suggest decreasing surface supersaturation. An alternative explanation could be offered based on the Al-adatom migration length. If critical supersaturation for island formation is reached on the surface terraces, the decreased migration length reduces the island capture  and thus growth rate, analogous to step-bunching. However, it is difficult to estimate the dependence surface supersaturation on V/III-ratio due to various process parameters. Further investigation on the relation between V/III-ratio and supersaturation is necessary.

In comparison with the 4$^{\circ}$ miscut substrates, no step bunching is observed in a micron scale for the AlN layers grown on 1$^{\circ}$ miscut substrate. The behavior could be attributed to wider surface terraces due to the decreased miscut. Decreasing the miscut angle has been shown to increase the parameter window for step-flow growth as the change between the dominating growth mode is less abrupt \cite{transition}. If the diffusion length of the ad-atoms is greater than the surface terrace width, step bunching morphology becomes more probable. Al diffusion length of 31~nm has been estimated for Al-polar AlN while N-polar AlN exhibits a shorter diffusion lenght \cite{AlNdiffusion,Diffusion}. By assuming that the two-bilayer-high steps formed on the SiC surfaces during the $H_{2}$ surface cleaning are uniformly distributed, the average terrace width is 28.9~nm and 7.24~nm for 1$^{\circ}$ and 4$^{\circ}$ miscut 4H-SiC substrate, respectively \cite{SiCsteps,SiCH2etch}. 

\begin{figure}[htb]
\begin{center}
\includegraphics[width=0.8\linewidth]{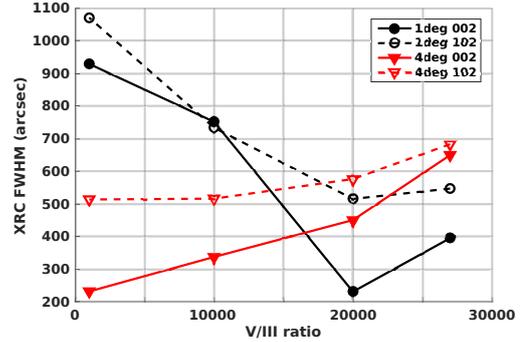}
\caption{XRC FWHMs of symmetrical 002 (filled symbols) and skew-symmetrical 102 (open symbols) scans for 200-nm-thick N-polar AlN layers grown on 4H-SiC substrates with 1$^{\circ}$ miscut (black) and 4$^{\circ}$ miscut (red) at 1150$^{\circ}C$ nominal substrate surface temperature with varying V/III-ratio.}
\label{XRD1440}
\end{center}
\end{figure}

Figure \ref{XRD1440} presents X-ray rocking curves (XRC) around symmetrical (002) and skew-symmetrical (102) reflections for the V/III-ratio series. In III-Ns, screw dislocations contribute to the broadening of the (002) peak, while both screw and edge dislocations contribute to the broadening of the (102) peak \cite{xrayIIIV}. Based on Figure \ref{XRD1440}, the crystalline quality of the N-polar AlN layers strongly depends on the V/III-ratio. The AlN layers grown on the 4$^{\circ}$ and 1$^{\circ}$ miscut substrates had vastly different optimum V/III-ratios which were 1000 and 20000, respectively. The smallest FWHMs were measured from the AlN layer grown on 4$^{\circ}$ miscut substrates with V/III ratio of 1000 were 232 arcsec and 514 arcsec for 
the (002) and the (102) scan, respectively. The FWHMs increase with increasing the V/III ratio, suggesting degrading crystalline quality. Based on our previous study we believe this to be the optimum V/III ratio since hexagonal hillocks were formed when V/III ratio was decreased below 1000 \cite{JoriPolarity}. We consider that the step flow growth mode contributes to high crystalline quality. 
On the other hand, for the AlN layers grown on the 1$^{\circ}$ miscut substrates the V/III-ratio of 20000 resulted in the smallest FWHMs of the (002) and the (102) scans which were 231 arcsec and 515 arcsec, respectively. The higher V/III ratios than 20000 resulted in larger FWHMs, suggesting degrading crystalline quality. Somewhat analogous to the 4$^{\circ}$ miscut substrate, highest crystalline quality is achieved at the lowest V/III-ratio sufficient to suppress the formation of the hexagonal hillocks. 

The highest reported crystalline quality in N-polar AlN has been achieved with molecular beam epitaxy. A 200-nm-thick N-polar AlN layer was coherently grown on SiC at very low growth temperature exhibiting XRC FWHMs of 120 arcsec and 210 arcsec for (002) and (102) scans \cite{HiroMBE}. On the other hand, relaxed AlN grown on SiC by MOVPE, an AlN layer grown using V/III-ratio of 24000 showed small XRC FWHMs of 468 arcsec and 684 arcsec for (002) and (102) scans, respectively \cite{NpolarhighNH3}. Despite relaxation, we achieved further high crystalline quality in N-polar AlN MOVPE growth.

\begin{figure}[htb]
\begin{center}
\includegraphics[width=0.6\linewidth]{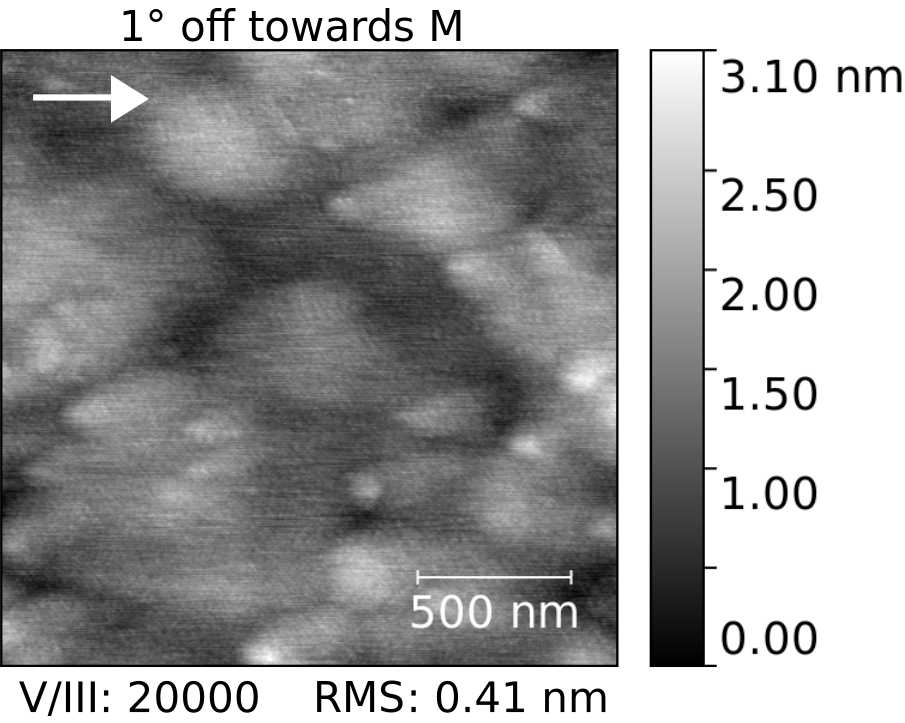}
\\
\includegraphics[width=0.6\linewidth]{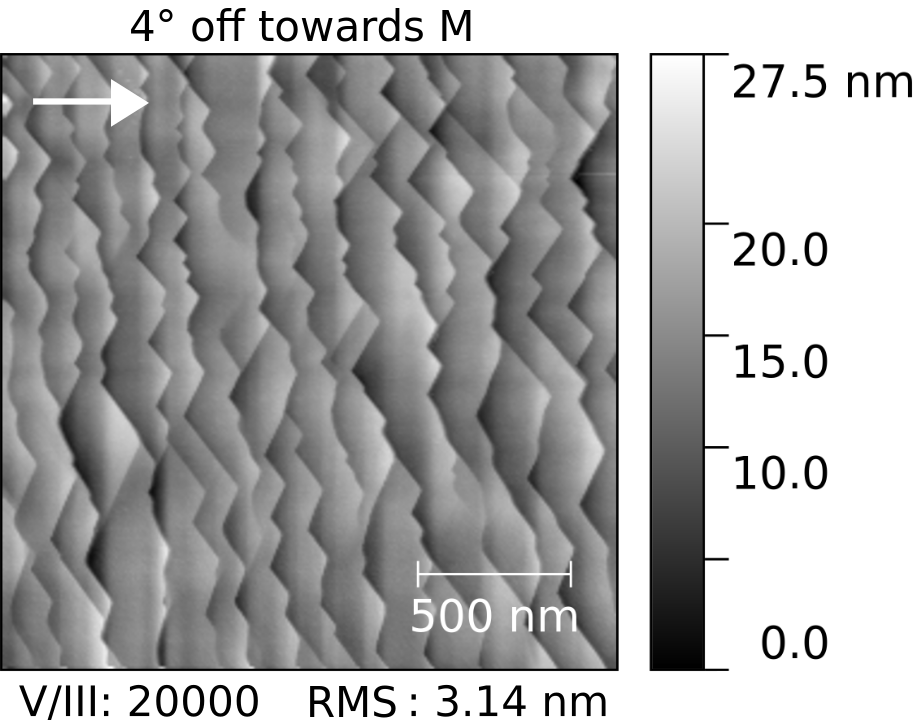}
\caption{2~$\mu m \times 2~\mu m$ AFM images of 200-nm-thick N-polar AlN layers grown on 4H-SiC 1$^{\circ}$ (top) and 4$^{\circ}$ (bottom) miscut substrates at 1150$^{\circ}C$ nominal substrate surface temperature with a V/III-ratio of 20000. The substrate miscut direction is indicated by white arrow.}
\label{AFM}
\end{center}
\end{figure}

Figure \ref{AFM} presents AFM $2~\mu m \times 2~\mu m$ images from the AlN layers grown on 1$^{\circ}$ (top) and 4$^{\circ}$ (bottom) miscut substrates with a V/III-ratio of 20000. The surface of the AlN layer grown on the 1$^{\circ}$ miscut subsrate is smooth with a root-mean-square (RMS) roughness of 0.41~nm. No step bunching or hexagonal hillocks are seen. In contrast, the AlN layer grown on the 4$^{\circ}$ miscut subsrate has a high RMS roughness of 3.14~nm due to step bunching, with an average step height of 9~nm. The steps propagate towards the substrate miscut of $<\bar{1}100>$, indicating step-flow growth mode. This indicates that it is possible to achieve simultaneous suppression of hexagonal hillocks and step bunching for N-polar AlN by controlling the Al-adatom migration with optimal growth conditions on low off-axis substrates. When compared to previous results, AFM 2~$\mu m \times 2~\mu m$ scan RMS roughness of 1.1~nm and 2~nm have been reported for 90-nm and 120-nm-thick N-polar AlN grown on SiC, respectively \cite{NpolarhighNH3,NpolarmedNH3_3}. In these studies, the large roughness of N-polar AlN surfaces were attributed to step bunching. N-polar III-Ns grown on substrates with miscut angle lower than 2$^{\circ}$ by MOVPE exhibited a high density of hexagonal hillocks \cite{miscut2stepbunch}. 160-nm-thick N-polar AlN on 0.2$^{\circ}$-off axis SiC grown by MBE exhibited 0.6~nm RMS roughness despite existence of hexagonal hillocks with a density of ~$10^{7}cm^{-2}$  \cite{HiroMBE}. 

\begin{figure}[htb]
\begin{center}
\includegraphics[width=0.6\linewidth]{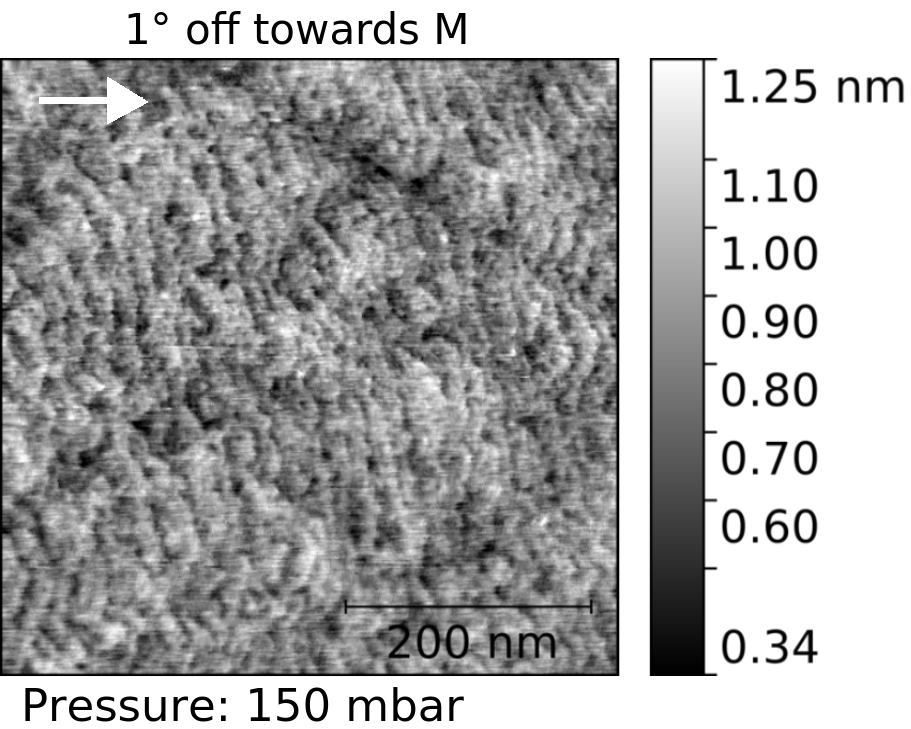}
\\
\includegraphics[width=0.6\linewidth]{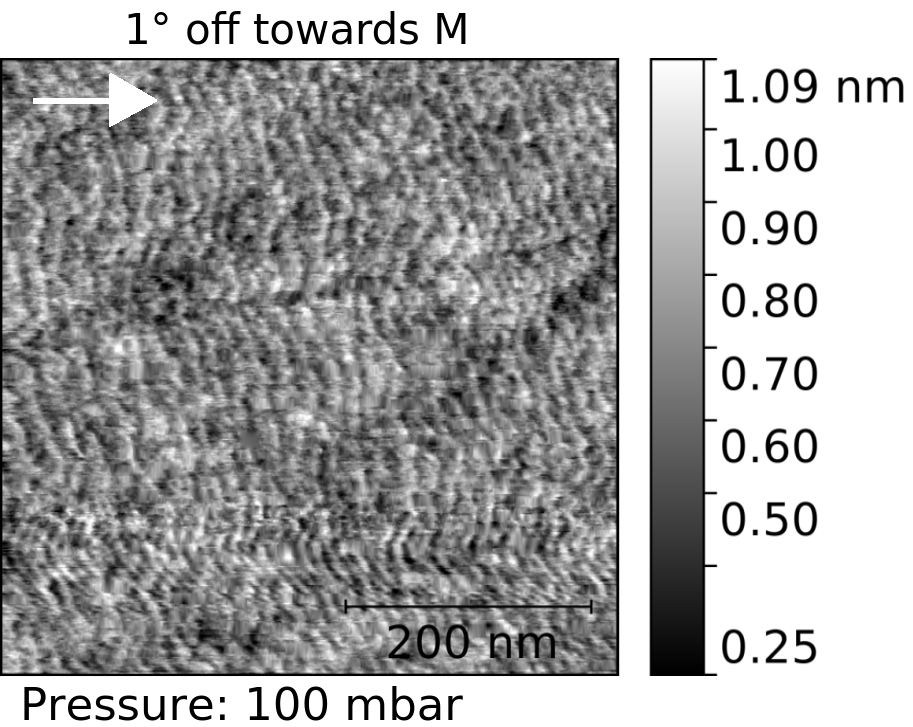}
\caption{$500~nm \times  500~nm$ AFM images of 4H-SiC substrates with 1$^{\circ}$ miscut after 20~min cleaning in $H_{2}$ atmosphere using 150~mbar pressure (top) and 100~mbar pressure (bottom) at 1180$^{\circ}C$ nominal substrate surface temperature. The substrate miscut direction is indicated by white arrow.}
\label{Bake}
\end{center}
\end{figure}

The substrate preparation prior to growth can have a significant impact on the crystalline quality of the epitaxial layer. The substrate $H_{2}$ surface cleaning used in the V/III-series was originally optimized for the 4$^{\circ}$ miscut substrate. A pressure of 100~mbar was then used during the substrate cleaning. Changing the cleaning conditions can effect the surface profile significantly \cite{SiCsteps,SiCH2etch}. Figure \ref{Bake} presents $500~nm \times 500~nm$ AFM images of 4H-SiC substrates after 20~min cleaning in $H_{2}$ atmosphere using 150~mbar pressure (top) and 100~mbar pressure (bottom) at 1180$^{\circ}C$ nominal substrate surface temperature. It can be seen that two-bilayer-high steps have formed on the substrate surfaces. The average calculated terrace width was 20~nm and 18~nm for 150~mbar and 100~mbar cleaning pressures, respectively, roughly corresponding to the previously calculated estimate of 28.9~nm. The difference could be due to a slight variation between the nominal and actual miscut angle. Based on Figure \ref{Bake} the 100~mbar cleaning pressure creates a well defined step-and-terrace structure whereas the substrate cleaned in 150~mbar pressure shows a more meandering step profile.

\begin{figure}[htb]
\begin{center}
\includegraphics[width=0.6\linewidth]{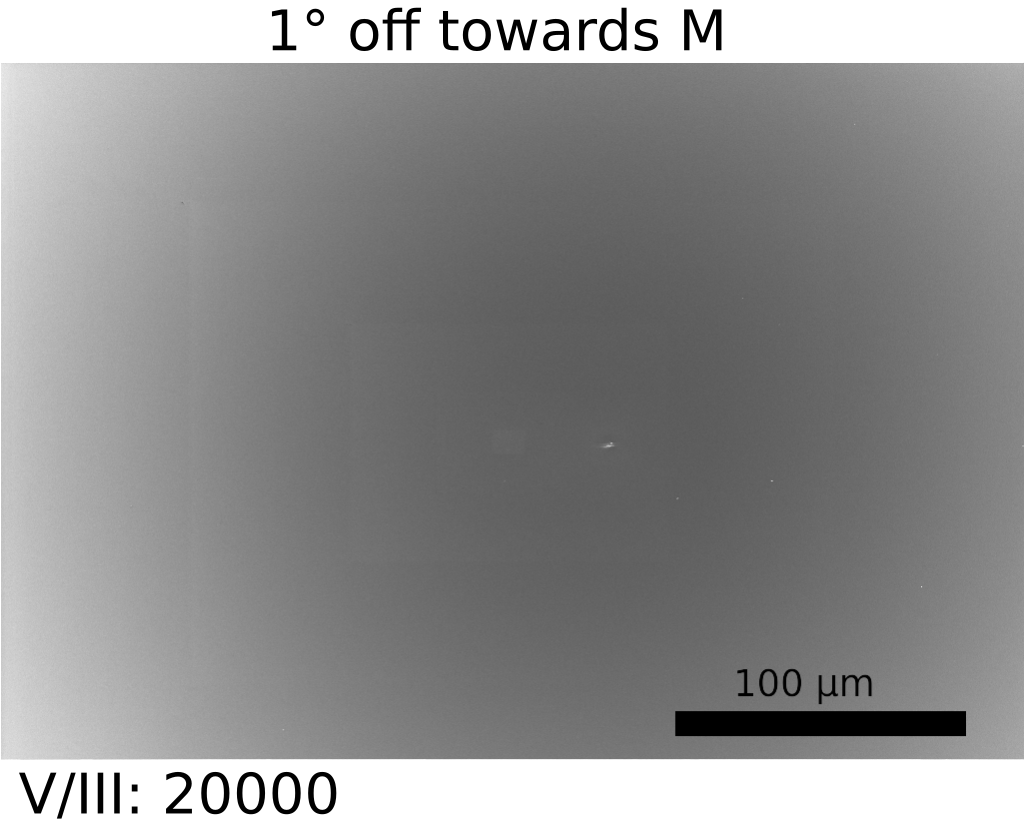}
\caption{Large area SEM image of a 200-nm-thick N-polar AlN layer grown on 4H-SiC 1$^{\circ}$ miscut substrate at 1150$^{\circ}C$ nominal substrate surface temperature with a V/III-ratio of 20000 and optimal substrate cleaning pressure of 100~mbar. A particle can be seen on the AlN layer surface in the center of the image.}
\label{LargeSEM}
\end{center}
\end{figure}

Using 100~mbar cleaning pressure, 200-nm-thick AlN layers were grown on 1$^{\circ}$ and 4$^{\circ}$ miscut substrates using a V/III-ratio of 20000. Figure \ref{LargeSEM} presents a large area SEM image of the AlN layer grown on the 1$^{\circ}$ miscut substrate. No hexagonal hillocks or other defects were discovered. 
\begin{figure}[htb]
\begin{center}
\includegraphics[width=0.6\linewidth]{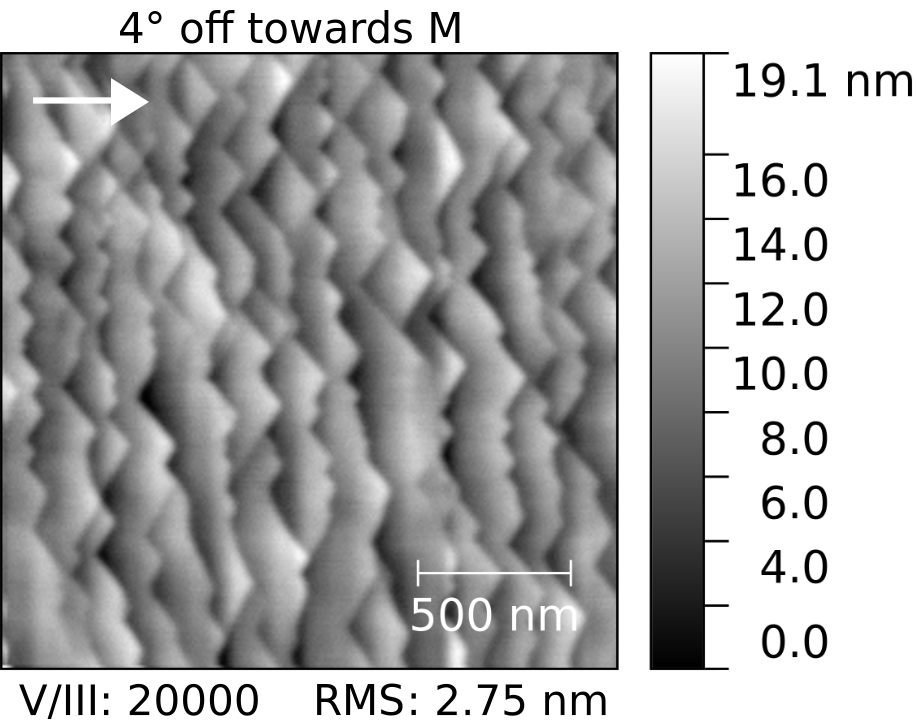}
\\
\includegraphics[width=0.6\linewidth]{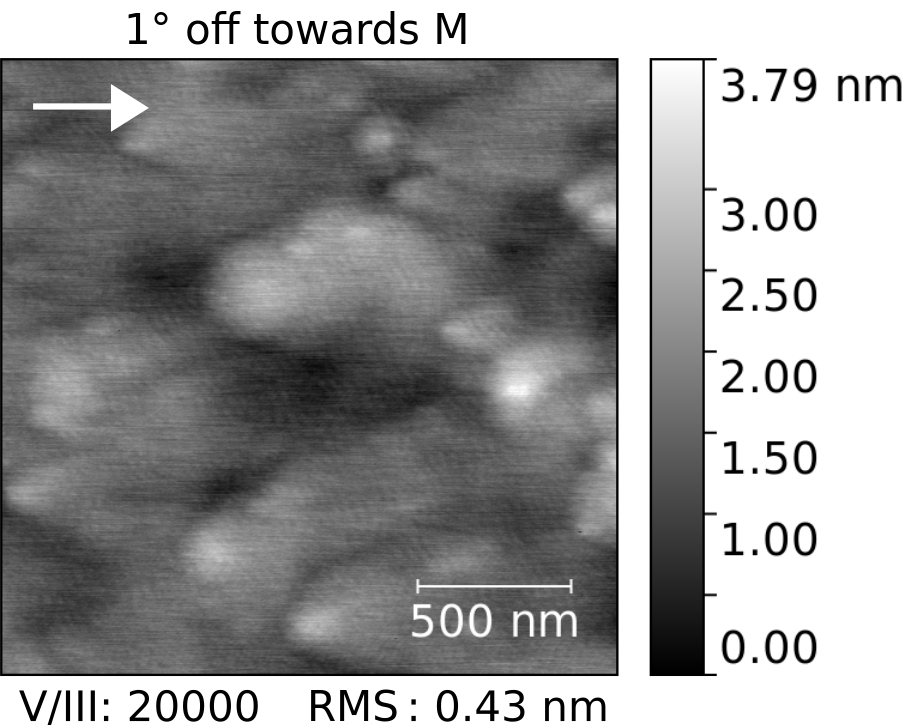}
\caption{2~$\mu m \times  2~\mu m$ AFM images of 200-nm-thick N-polar AlN layers grown on 4H-SiC 1$^{\circ}$ (top) and 4$^{\circ}$ (bottom) miscut substrates at 1150$^{\circ}C$ nominal substrate surface temperature with a V/III-ratio of 20000 and substrate cleaning pressure of 100~mbar. The substrate miscut direction is indicated by white arrow.}
\label{AFM2}
\end{center}
\end{figure}
Figure \ref{AFM2} presents 2~$\mu m \times 2~\mu m$ AFM images from the AlN layers grown on 1$^{\circ}$ (top) and 4$^{\circ}$ (bottom) miscut substrates. Compared to the AlN layers in Figure \ref{AFM}, changing the substrate cleaning pressure has no significant effect on the layer morphology or RMS roughness. The RMS roughnesses were 0.43~nm and 2.75~nm for AlN layers grown on 1$^{\circ}$ and 4$^{\circ}$ miscut substrate, respectively. The crystalline quality of AlN layers grown on the 1$^{\circ}$ miscut substrates was improved. XRC FWHMs of 203 and 389 arcsec for (002) and (102) reflections were measured, from a sample grown on the 1$^{\circ}$ miscut substrate using the optimal V/III-ratio of 20000 together with substrate cleaning pressure of 100~mbar. We achieved the highest crystalline-quality N-polar AlN layers grown by MOVPE.

%

\section{Conclusions}

200-nm-thick N-polar AlN layers were grown by MOVPE on C-surface of SiC-4H substrates with an intentional miscut of 1$^{\circ}$ or 4$^{\circ}$ towards $<\bar{1}100>$. The optimal V/III-ratios for 1$^{\circ}$ and 4$^{\circ}$ miscut substrates were found to be 20000 and 1000, respectively. MOVPE grown N-polar AlN layer without hexagonal hillocks or step bunching was achieved using a 4H-SiC substrate with an intentional miscut of 1$^{\circ}$. The XRC FWHMs were 203 arcsec and 389 arcsec for (002) and (102) reflections, respectively. The RMS surface roughness was 0.43~nm for AFM 2~$\mu m \times 2~\mu m$ scan. We found that the high V/III ratio and small off angle of substrates contributes to high crystalline quality in high-temperature N-polar AlN growth.

\section*{Acknowledgements}

This work was supported by the Academy of Finland (grant 297916), the Foundation for Aalto University Science and Technology and JSPS KAKENHI Grant No. 16H06424 and 17K14110. The research was performed at the OtaNano - Micronova Nanofabrication Centre of Aalto University.

\section*{References}

\bibliographystyle{elsarticle-num}
\bibliography{AlN}

\end{document}